\documentclass[english]{IEEEtran}
\usepackage[T1]{fontenc}
\usepackage[latin9]{inputenc}
\usepackage{verbatim}
\usepackage{booktabs}
\usepackage{graphicx}

\makeatletter

\providecommand{\tabularnewline}{\\}

\makeatother

\usepackage{babel}
\begin{document}

\title{Classification of Radio Signals and \\
HF Transmission Modes with Deep Learning}

\author{Stefan Scholl \\
dc9st@panoradio-sdr.de}
\maketitle
\begin{abstract}
This paper investigates deep neural networks for radio signal classification.
Instead of performing modulation recognition and combining it with
further analysis methods, the classifier operates directly on the
IQ data of the signals and outputs the transmission mode. A data set
of radio signals of 18 different modes, that commonly occur in the
HF radio band, is presented and used as a showcase example. The data
set considers HF channel properties and is used to train four different
deep neural network architectures. The results of the best networks
show an excellent accuracy of up to 98\%.
\end{abstract}

\section{Introduction}

Classification of radio signals is an essential task in signal intelligence
and surveillance applications and is recently adopted in applications
like cognitive radio and dynamic spectrum access to continuously monitor
the spectrum and its occupancy by various different radio signals,
modes and services.

Traditional approaches to radio signal classification rely on signal
features based on probabilistic methods, statistics or cyclostationarity.
These features need to be carefully designed by expert developers
and thus depend on their experience and knowledge on the problem structure
\cite{Zhu:2015:AMC:2800343}.

Recently advanced machine learning techniques, like deep neural networks,
gained huge interest and showed extremely good performance for classification
tasks in various applications. Instead of formulating hand-crafted
features by a designer, a training algorithm uses large amounts of
labelled example data to learn and extract good features from the
data that are discriminative for the classification task. While this
data-driven approach first proved to be very successful for object
classification in image processing \cite{Krizhevsky:2012:ICD:2999134.2999257},
it quickly advanced to numerous further applications, such as different
challenges in radio communications \cite{OShea2017a}. 

For the task of signal classification, the neural network trains on
large amounts of raw data of radio signals (e.g. IQ data samples),
such that it is enabled to distinguish the different signal classes.
Especially for the task of modulation classification, neural networks
have shown to exhibit very good performance \cite{OShea2017,Liu2017}.

Although recent work is based on IQ data samples as input, the output
classes are restricted to modulation types (e.g. FSK, PSK, AM). To
determine the actual transmission mode or service (Wi-fi, mobile telephony
(GSM, LTE), digital radio (DAB), radioteletype (RTTY), AM broadcasting,
etc.) further processing is required, like the additional measurement
of baud rate, pulse shape or bit pattern structures. These additional
features are not learnt by the neural network trained on modulation
types and therefore this second step follows the classical way of
designing expert features by hand.

This paper investigates how neural networks can be used to classify
signals by their transmission mode directly, instead of classifying
only modulation types. The approach purely follows the data driven
paradigm by mapping input IQ data to the output modes directly. As
an example application this paper considers radio signals typically
present in the HF band (3-30 MHz), because this wireless band contains
many different modes that coexist closely spaced in the frequency
spectrum. In total, 18 different HF transmission modes are considered
for classification. Four different types of neural networks are trained
on a synthetically generated data set, considering a noisy HF channel
environment under imperfect receiver conditions.

The investigation shows, that neural networks are very powerful in
classifying signals into their transmission modes even if they exhibit
very similar properties. An accuracy of 98\% for moderate SNRs can
be obtained with reasonable training effort. 

The remainder of the paper is structured as follows: Section \ref{sec:Data-Set}
describes the training data in more detail, Section \ref{sec:Models-and-Training}
introduces the models and the training process, followed by the results
in Section \ref{sec:Results}.

\section{\label{sec:Data-Set}Data Set}

\subsection{Modes and Data Format}

The training data consists of the 18 different transmission modes
shown in Table \ref{tab:Transmission-Modes}, of which many are commonly
found in the HF bands. This includes AM broadcasting, single-sideband
(SSB) audio, radioteletype (RTTY, Navtex) \cite{itu_navtex}, morse
code \cite{itu_morse}, facsimile and further modes for digital data
transmission like PSK31 \cite{itu_psk}, Olivia \cite{itu_olivia}
and others. The selected 18 modes cover very different modulation
techniques, including analog (AM, facsimilie and SSB) and various
digital modulation types (PSK, FSK, M-FSK, OOK, multicarrier). However,
there are also very similar modes, such as RTTY45 and RTTY50, that
differ only by their baud rate of 45 and 50 Bd and are expected to
be especially hard to classify. Other similar modes are PSK31 and
PSK63, as well as SSB in upper (USB) and lower (LSB) sideband.

\begin{table}[h]
\begin{centering}
\begin{tabular}{lll}
\toprule 
Mode Name & Modulation & Baud Rate\tabularnewline
\midrule
Morse Code  & OOK & variable\tabularnewline
PSK31  & PSK & 31\tabularnewline
PSK63  & PSK & 63\tabularnewline
QPSK31  & QPSK & 31\tabularnewline
RTTY 45/170 & FSK, 170 Hz shift & 45\tabularnewline
RTTY 50/170 & FSK, 170 Hz shift & 50\tabularnewline
RTTY 100/850 & FSK, 850 Hz shift & 850\tabularnewline
Olivia 8/250  & 8-MFSK & 31\tabularnewline
Olivia 16/500 & 16-MFSK & 31\tabularnewline
Olivia 16/1000 & 16-MFSK & 62\tabularnewline
Olivia 32/1000  & 32-MFSK & 31\tabularnewline
DominoEx & 18-MFSK & 11\tabularnewline
MT63 / 1000 & multi-carrier & 10\tabularnewline
Navtex / Sitor-B & FSK, 170 Hz shift & 100\tabularnewline
Single-Sideband (upper) & USB & -\tabularnewline
Single-Sideband (lower) & LSB & -\tabularnewline
AM broadcast & AM & -\tabularnewline
HF/radiofax & radiofax & -\tabularnewline
\bottomrule
\end{tabular}
\par\end{centering}

\medskip{}

\caption{\label{tab:Transmission-Modes}Transmission Modes }
\end{table}

The training data is generated synthetically by simulation. The raw
data used to generate the training signals is plain text for the digital
modes and speech and music (various genres) for the analog modes.
Fax data is generated by modulating different black \& white images,
such as e.g. transmitted by weather services. Then standard software
modulates the raw signals in order to obtain baseband signals, which
are then artificially distorted by a HF channel model, described in
detail below. 

Transmission in the HF band is mostly characterized by comparably
small bandwidths (often less than 3 kHz) and therefore low data rates.
The data set contains vectors of complex IQ data of length 2048 with
a sample rate of 6 kHz. Thus a data vector corresponds to a time of
approximately 0.3~s. In total, the data set consists of 120,000 training
vectors and another 30,000 vectors for validation.

\subsection{HF Channel Properties}

Since signals that are transmitted via the HF band exhibit special
distortions, these effects needs to be reflected by the training data.
The HF band ranges from 3 to 30 MHz and is characterized by sky wave
propagation, i.e. radio waves do not travel by line-of-sight, but
are reflected by the earth's ionosphere, which enables intercontinental
communication with little technical effort. In fact, different ionospheric
layers, that may be located in different heights, contribute to the
propagation of radio waves. This complex propagation behaviour presents
a multi path propagation environment resulting in fading effects.
Moreover, these ionospheric layers are not stationary, but are in
continuous movement, which introduces varying doppler shifts.

For modelling radio wave propagation in the HF band, the Watterson
model is commonly applied \cite{Watterson1970}, that covers fading
and doppler shift introduced by multipath propagation effects. The
ITU has defined several channel models based on the Watterson model,
called CCIR 520 \cite{CCIR_520}. CCIR 520 includes different scenarios
for complex wave propagation, that differ in the amount of distortion
introduced, i.e. frequency spread, frequency offset and differential
time delay. The scenarios employed in the training data sets are \emph{good
conditions}, \emph{moderate conditions}, \emph{bad conditions}, \emph{flutter
fading} and \emph{doppler fading}, plus data vectors without any fading
and doppler distortion.

In addition, all data vectors are distorted by Gaussian noise (AWGN)
of different strength, such that the SNR of the data is evenly distributed
from -10 to +25 dB. To account for non-coherent reception and receiver
frequency mismatch also random phase and frequency offsets are applied.
In summary, the training data set incorporates the following types
of distortion to provide robust training results that perform well
in real-world scenarios:
\begin{itemize}
\item CCIR 520 channel models 
\item AWGN noise (from -10 to +25 dB) 
\item random frequency offset (+- 250 Hz) 
\item random phase offset
\end{itemize}

\section{\label{sec:Models-and-Training}Models and Training}

\begin{figure}[h]
\begin{centering}
\includegraphics[scale=0.47]{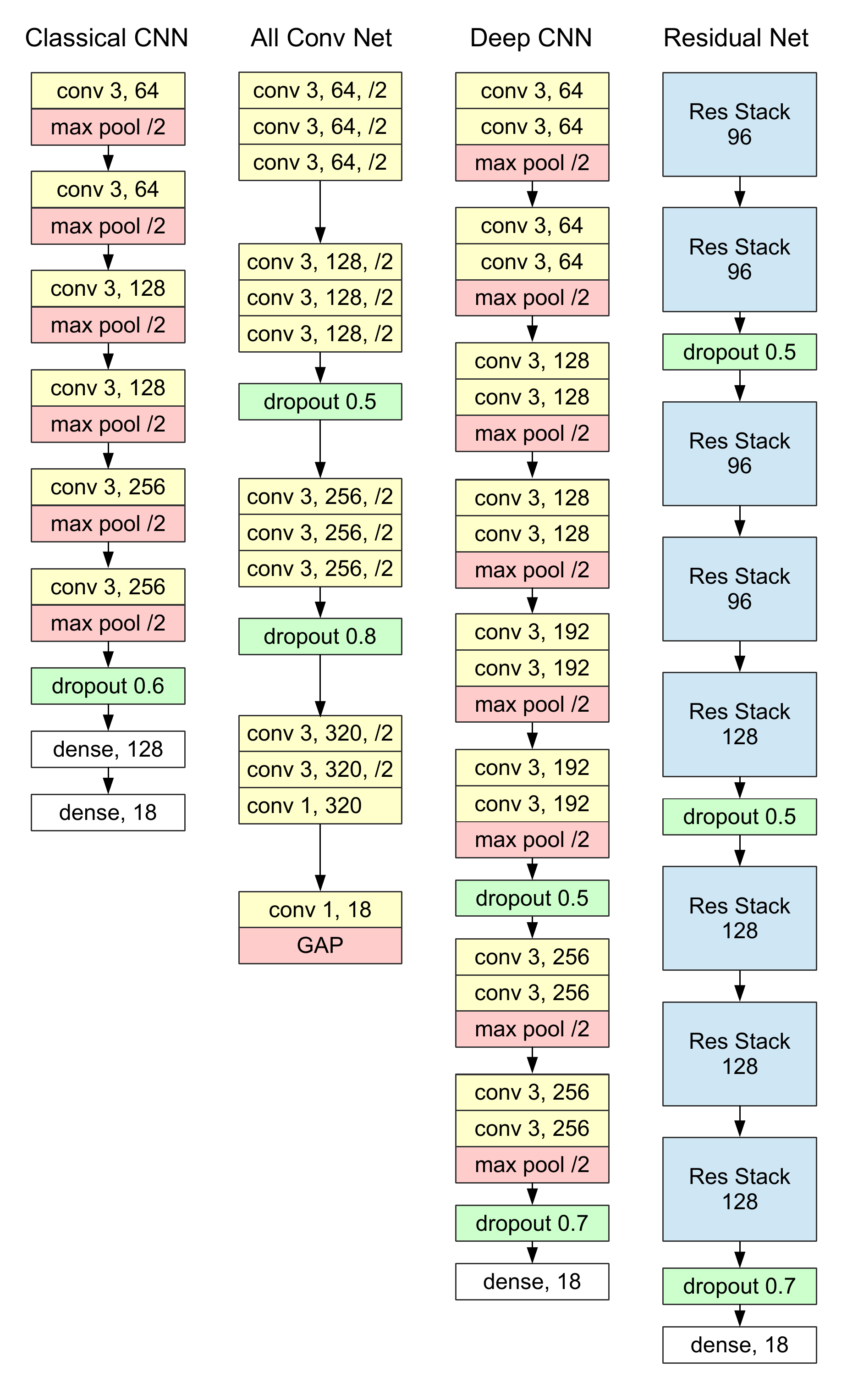}
\par\end{centering}

\caption{\label{fig:Different-neural-network}Different neural network architectures}
\end{figure}

\begin{figure}[h]
\begin{centering}
\includegraphics[scale=0.5]{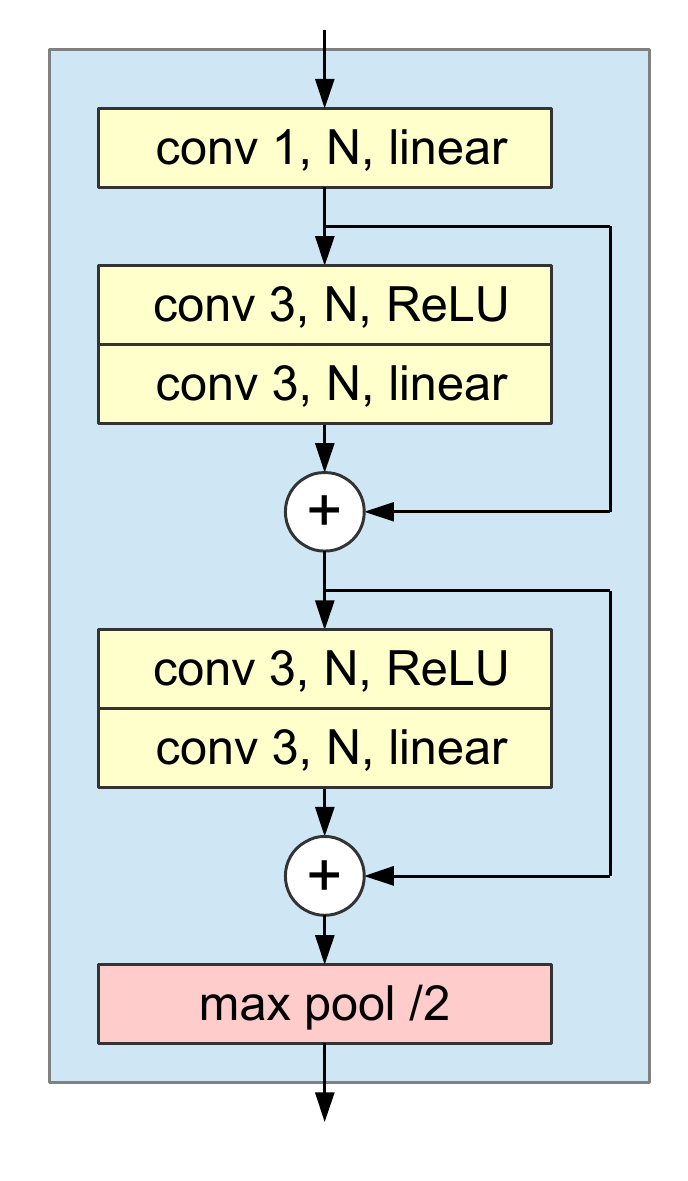}
\par\end{centering}

\caption{\label{fig:The-residual-stack}The residual stack as used in the residual
net. The parameter N provides the number of filters for the stack. }
\end{figure}

This paper investigates four neural networks or models for signal
classification, that range from ordinary convolutional neural nets
(CNN) \cite{lecun-bengio-95a} to advanced models, like residual nets
\cite{He2015}. The structure of the models is depicted in Fig. \ref{fig:Different-neural-network}.
\begin{itemize}
\item \textbf{Classical CNN}: It consists of six convolutional layers each
followed by max pooling. At the output two large dense layers are
located.
\item \textbf{All Convolutional Net}: It is only composed of convolutional
layers with stride 2 (instead of max pooling layers) and it has a
global average pooling layer at its output \cite{Springenberg2014}.
\item \textbf{Deep CNN}: It uses 16 convolutional layers, with max pooling
after every second convolutional layer and a single small dense layer
at the output, roughly following the concept of the VGG net \cite{Simonyan2014}.
\item \textbf{Residual Net}: It uses a large number of layers enhanced by
residual connections. While residual nets as described in \cite{He2015}
(including the bottleneck architecture) did not provide convincing
results, an arrangement of eight 5-layer residual stacks of \cite{OShea2017}
proved to be more appropriate for the task of signal classification
and is applied here (Fig. \ref{fig:The-residual-stack}).
\end{itemize}
The nets are chosen to have a similar number of parameters around
1.4M. They adopt ReLU activation functions and softmax for the output
layers. The nets use dropout layers for regularization and batch normalization.
Following the ideas of \cite{Simonyan2014} the convolutional layers
mostly use a filter size of 3.

Training is done with adam optimization \cite{Kingma2014} enhanced
by a learning rate scheduler, that reduces the learning rate as soon
as the training process plateaus. The batch size is 128 and the number
of training epochs is 30 to keep the training time moderate while
achieving very good training accuracy.

\section{\label{sec:Results}Results}

Table \ref{tab:model-overview-results} shows a comparison of the
results for the four models that have been trained with the previously
described data set. The validation data set to measure accuracy also
contains signal data evenly distributed over the whole SNR range from
-10 to 25 dB. Therefore the provided accuracy values in Table \ref{tab:model-overview-results}
are an average over all SNR values.

Best performance show the deep CNN (17 layers) with 93.7\% and the
residual net (41 layers) with 94.1\%. Although the residual net has
much more layers, it provides only a minor improvement in accuracy.
Larger improvement in accuracy by the much deeper residual net may
be obtained when using much more training time than the 30 epochs
used for economic training in this paper.

\begin{table}[h]
\begin{centering}
\begin{tabular}{lrccc}
\toprule 
Model & Layers & \#Parameters & Accuracy & Training Time\tabularnewline
\midrule
Classical CNN & 8 & 1.4 M & 85.8\% & 1.2 h\tabularnewline
All conv net & 13 & 1.3 M & 90.3\% & 0.7 h\tabularnewline
Deep CNN & 17 & 1.4 M & 93.7\% & 1.9 h\tabularnewline
Residual net & 41 & 1.4 M & 94.1\% & 5.3 h\tabularnewline
\bottomrule
\end{tabular}
\par\end{centering}

\medskip{}

\caption{\label{tab:model-overview-results}Overview of the different models
and their performance}
\end{table}

\begin{figure}[h]
\begin{centering}
\includegraphics[scale=0.4]{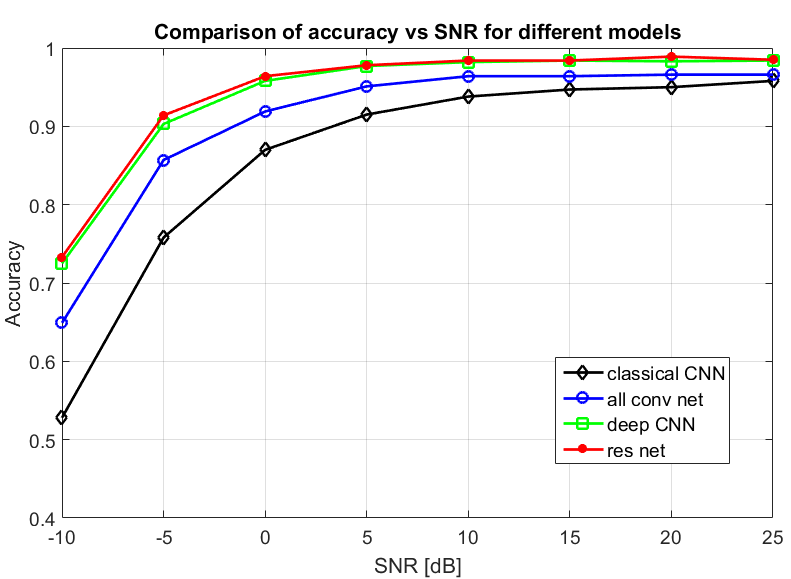}
\par\end{centering}

\caption{\label{fig:acc_over_snr}Comparison of the results of the different
models}
\end{figure}

Fig. \ref{fig:acc_over_snr} shows the accuracy of the different models
over SNR. Even for small SNR values of -5 dB, the accuracy for the
best models is above 90\%. When SNR is above 5 dB the accuracy increases
to an excellent value of approximately 98\%.

\begin{figure}[h]
\begin{centering}
\includegraphics[scale=0.2]{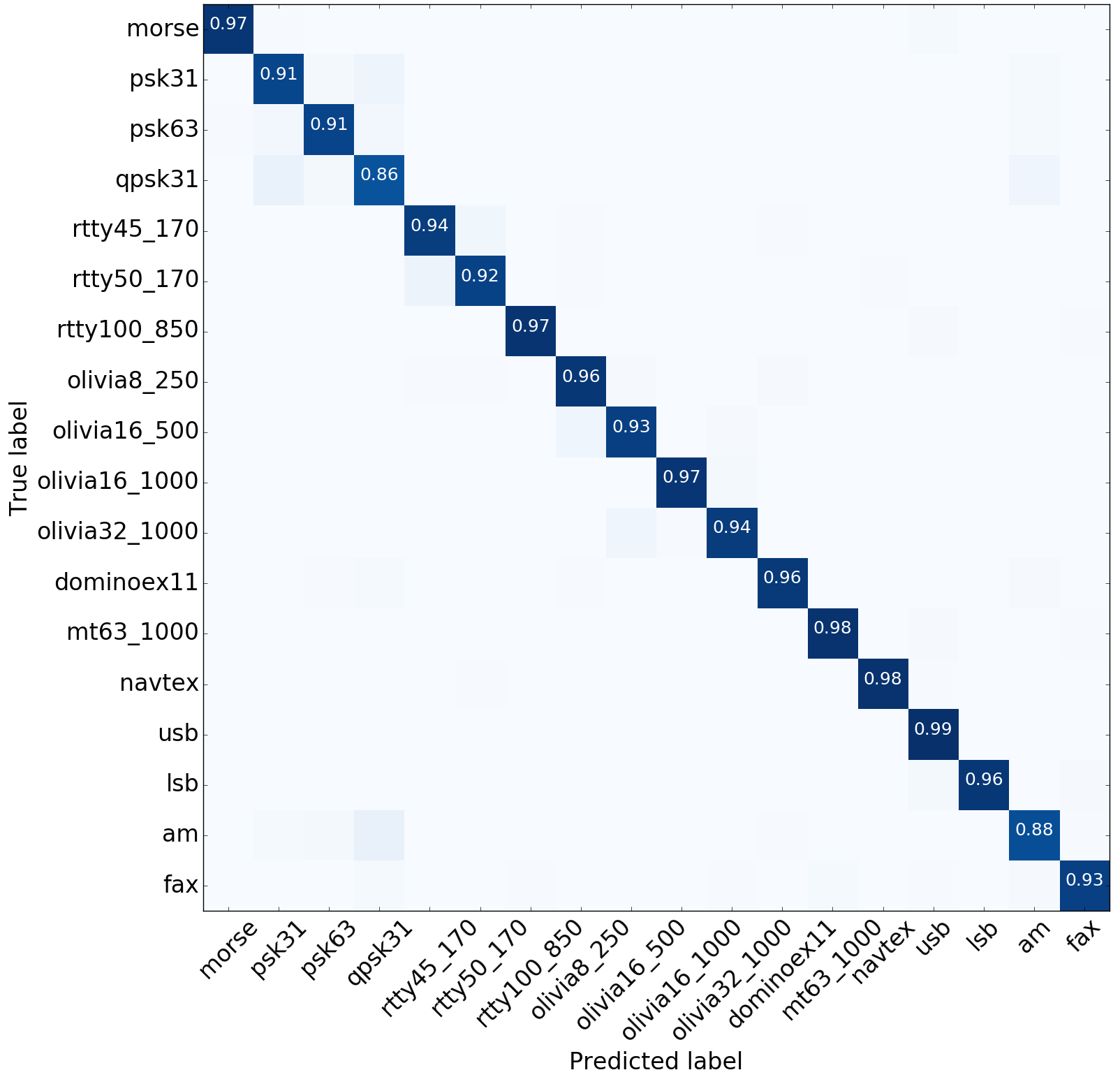}
\par\end{centering}

\centering{}\caption{\label{fig:Confusion-Matrix}Confusion matrix for the residual net
(all SNR values)}
\end{figure}

Fig. \ref{fig:Confusion-Matrix} shows the confusion matrix for the
residual net, that provides the best results. Since the overall performance
of the net is very good, confusions are very rare. Minor confusions
occur between QPSK and BPSK modes. Also RTTY45 and RTTY50 tend to
rarely be confused, because the only difference between these modes
is the slightly deviating baud rate of 45 and 50. Although LSB und
USB are quite similar modes, they are classified with very high reliability.

\section{Conclusion}

This paper presented four different neural networks for the classification
of radio signals. Instead of focusing on modulation recognition, the
models learn to classify different transmission modes directly. This
saves additional post-processing required to determine the modes from
modulation and other signal parameters. An exemplary data set with
18 different transmission modes that occur the HF band has been utilized,
with an excellent accuracy of up to 98\%.

\bibliographystyle{IEEEtran}
\bibliography{signal_classification}

\end{document}